\documentclass[Conference,a4paper]{IEEEtran}
\IEEEoverridecommandlockouts
\usepackage{color}
\usepackage{graphicx}
\usepackage{epstopdf}
\usepackage{amsmath}
\usepackage{amssymb}
\usepackage{algorithm}
\usepackage{algorithmic}
\usepackage{amsmath}
\usepackage{multirow}
\usepackage{booktabs}
\usepackage{array}
\usepackage{amsthm}
\usepackage{stfloats}
\usepackage{caption}
\usepackage{subfigure}
\usepackage{bm}
\usepackage{booktabs}
\usepackage{setspace}

\newcommand{\be}{\begin{equation}}
\newcommand{\ee}{\end{equation}}
\newcommand{\bea}{\begin{eqnarray}}
\newcommand{\eea}{\end{eqnarray}}
\newcommand{\ba}{\begin{array}}
\newcommand{\ea}{\end{array}}

\title{DRL-based Joint Beamforming and BS-RIS-UE Association Design for RIS-Assisted mmWave Networks
\thanks{This work is supported in part by the National Natural Science Foundation of China (Grant No. 61971088, 62071083, 62071105, U1808206, and U1908214), the Natural Science Foundation of Liaoning Province (Grant No. 2020-MS-108), in part by the Fundamental Research Funds for the Central Universities (Grant No. DUT20GJ214, DUT21GJ208 and DUT20RC(3)029), in part by Dalian Science and Technology Innovation Project (Grant No. 2020JJ25CY001),  and  in part by the Open Research Fund of National Mobile Communications Research Laboratory, Southeast University (Grant No. 2021D08).}}

\author{\IEEEauthorblockN{Yuqian Zhu$^{\dag}$, Ming Li$^{\dag}$$^{\ddag}$, Yang Liu$^{\dag}$, Qian Liu$^{\dag}$, Zheng Chang$^{*}$, and Yulin Hu$^{\sharp}$
\vspace{-0.0 cm} }\\ 
\IEEEauthorblockA{$^{\dag}$ Dalian University of Technology, Dalian, Liaoning 116024, China \\ E-mail: \texttt{yqzhu@mail.dlut.edu.cn, \{mli,yangliu\_613,qianliu\}@dlut.edu.cn} } \\
\IEEEauthorblockA{$^{\ddag}$ National Mobile Communications Research Laboratory Southeast University, Nanjing, Jiangsu 210096, China} \\
\IEEEauthorblockA{$^{*}$ University of Electronic Science and Technology of China, Chengdu, Sichuan 611731, China \\ E-mail: \texttt{zheng.chang@jyu.fi} }\\
\IEEEauthorblockA{$^{\sharp}$ Wuhan University, Wuhan, Hubei 430072, China \\ E-mail: \texttt{yulin.hu@whu.edu.cn} }
}
\begin{document}
\maketitle
\pagestyle{empty}
\thispagestyle{empty}
\begin{abstract}
Reconfigurable intelligent surface (RIS)  is considered as an extraordinarily promising technology to solve the blockage problem of millimeter wave (mmWave) communications owing to its capable of establishing a reconfigurable wireless propagation. In this paper,  we focus on a RIS-assisted mmWave communication network consisting of multiple base stations (BSs) serving a set of user equipments (UEs). Considering the BS-RIS-UE association problem which determines that the RIS should assist which BS and UEs,
 we joint optimize BS-RIS-UE association and passive beamforming at RIS to maximize the sum-rate of the system. To solve this intractable non-convex problem, we propose a soft actor-critic (SAC) deep reinforcement learning (DRL)-based joint beamforming and BS-RIS-UE association design algorithm, which can learn the best policy by interacting with the environment using less prior information and avoid falling into the local optimal solution by incorporating with the maximization of policy information entropy. The simulation results demonstrate that the proposed SAC-DRL algorithm can achieve significant performance gains compared with benchmark schemes.

\end{abstract}
\begin{IEEEkeywords}
Reconfigurable intelligent surface, deep reinforcement learning, beamforming, BS-RIS-UE association, millimeter wave communications.
\end{IEEEkeywords}
\maketitle
\vspace{-0.3 cm}
\section{Introduction}
With the rapid growth of data transmission rate in future wireless applications, the fifth-generation (5G) communication technologies such as millimeter wave (mmWave) communications, massive multiple-input multiple-output (MIMO), and ultra-dense networks attempt to reach the explosive demand \cite{5Gsurvey}, \cite{5Gtechnology}. However, the mmWave communication is usually accompanied by the problems of small coverage, serious path loss, and susceptibility to obstruction of signals \cite{channel}.
Recently, the reconfigurable intelligent surface (RIS), which is a programmable meta-surface, has become a promising technology to overcome these challenges \cite{9326394}. When the line-of-sight (LOS) link in the mmWave system is blocked by obstacles, RIS can establish a new reflected link without consuming additional energy, which greatly improves the coverage of the base station (BS) and maintains the reliability of the mmWave communication systems.

Many researchers are committed to design RIS-assisted wireless communication systems with single BS \cite{yxh}-\cite{xy}.
However, as the density of BS deployment will be greatly increased in ultra-dense mmWave MIMO communication systems, each user equipment (UE) may be surrounded by multiple BSs. Therefore, user association plays a vital role in improving spectrum efficiency and system performance \cite{UA}.
The authors in \cite{MIRS} studied the optimal BS-RIS-UE association problem for maximizing the utility in a multi-RIS-aided wireless network.
 However, the traditional optimization-based algorithms will bring extremely high computational complexity to optimize the non-convex problem with multiple coupled variables across different BSs.

Fortunately, artificial intelligence (AI) techniques can efficiently solve massive data, mathematically difficult non-linear and non-convex problems. Particularly, deep reinforcement learning (DRL) seeks the optimal strategy through an agent-environment interactive learning, which is considered as an extremely promising candidate technology in the future communication network facing diversified demands. Recently, DRL-based methods have been used to solve intractable non-convex optimization problems in wireless communication systems \cite{CH}-\cite{FB}.
In \cite{CH}, a policy-based deep
deterministic policy gradient (DDPG) was developed to jointly design
the transmit beamforming matrix at BS and the phase-shift matrix at RIS.
In \cite{JG}, a distributed DRL-based dynamic downlink beamforming coordination method was proposed to adjust the beamformers of all BSs in the cellular network.
 In \cite{JK}, a dynamic control scheme based on multi-agent DRL was proposed to design UE powers, RIS beamformer, and BS combiners in an uplink multi-RIS-assisted multi-cell systems.
In \cite{FB},  a joint beamforming, power control, and interference coordination algorithm based on DRL was developed for a multi-access Orthogonal Frequency Division Multiplexing (OFDM) cellular network.
 While most exiting literature focused on joint design of active beamforming at BS and phase-shift at RIS, using DRL algorithm to solve the BS-RIS-UE association has not been investigated.

Motivated by the above analysis, in this paper, we utilize a soft actor-critic (SAC)-DRL algorithm to solve the joint beamformer and BS-RIS-UE association design problem.
Specifically, we consider a RIS-assisted mmWave network consisting of multiple BSs and aim to design a SAC-DRL based algorithm to jointly design passive beamforming at RIS and BS-RIS-UE association to maximize the sum-rate of the system.
 The proposed SAC-DRL algorithm can explore more stochastic policies and avoid falling into the local optimal solution by maximizing the reward and the policy information entropy.
The experimental results demonstrate that our proposed SAC-DRL algorithm can achieve better sum-rate performance with similar training episodes compared with benchmark schemes.
\vspace{-0.3 cm}
\section{System Model and Problem Formulation}
\subsection{System Model}
\begin{figure}[t]
\centering
\includegraphics[width = 2.8 in]{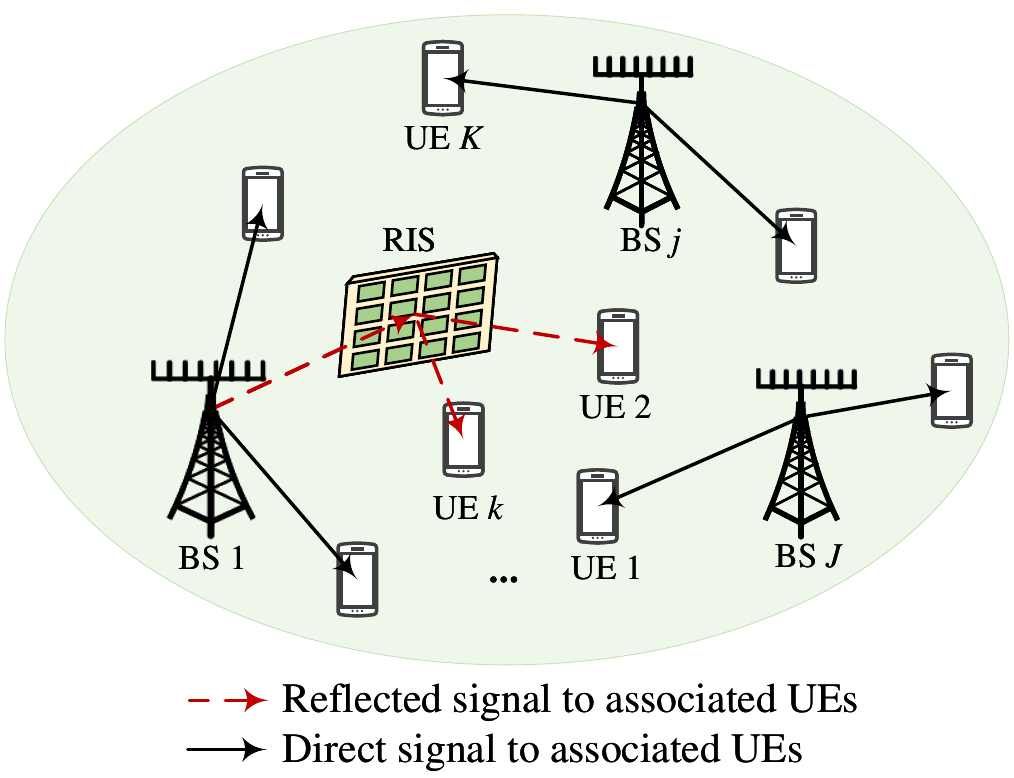}
\vspace{-0 cm}
\caption{A RIS-assisted multi-BS multi-UE MISO system.}
\label{fig:model}
\vspace{-0.4 cm}
\end{figure}

 As shown in Fig. \ref{fig:model}, we consider a RIS-assisted mmWave communication system consisting of $J$ BSs to serve $K$ UEs, where the sets of the BSs and  the UEs are denoted as $\mathcal{J} = \{1,2,\cdots,J\}$ and $\mathcal{K} = \{1,2,\cdots,K\}$, respectively. Each BS is equipped with $N>K$ transmit antennas, and each UE is equipped with merely one receive antenna.
  The RIS has $M$ reflection elements deployed on the surface of building and will be associated with one BS.
   We assume that the RIS only serves one BS during each coherent period. $\mathbf{G}_{j}\in\mathbb{C}^{M \times N}$, $\mathbf{h}_{\mathrm{r},j,k} \in \mathbb{C}^{M \times 1}$, $\mathbf{h}_{\mathrm{d},j,k} \in \mathbb{C}^{N \times 1}$, $j \in \mathcal{J}, k \in \mathcal{K}$, are equivalent channels from BS-$j$ to RIS, from RIS to UE-$k$, and from BS-$j$ to UE-$k$, respectively. Besides,
the passive beamforming of the RIS is denoted as $\mathbf{\Psi} = diag(\mathbf{f})$, where ${\mathbf{f}}={\mathbf{f}}_{\mathrm{h}} \otimes {\mathbf{f}}_{\mathrm{v}}$,
the elements of $ {\mathbf{f}}_{\mathrm{h}}$ and $ {\mathbf{f}}_{\mathrm{v}}$ can be represented by
\begin{equation}
\small
\begin{aligned}
&{\mathbf{f}}_{\mathrm{h}}({m}_{\mathrm{h}})=\frac{\exp \left(-j  \pi \cos \left(\phi\right) \sin \left(\theta\right)\left({m}_{\mathrm{h}}-1\right) \right)}{\sqrt{{M}_{\mathrm{h}}}}, \\
&{\mathbf{f}}_{\mathrm{v}}({m}_{\mathrm{v}})=\frac{\exp \left(-j \pi \sin \left(\phi\right) \left({m}_{\mathrm{v}}-1\right) \right)}{\sqrt{{M}_{\mathrm{v}}}},
\end{aligned}
\end{equation}
where ${m}_\mathrm{h}=1, \ldots, {M}_\mathrm{h}$ and ${m}_\mathrm{v}=1, \ldots, {M}_\mathrm{v}$ are the indices of RIS elements in horizontal and vertical directions, respectively. Also,
the $\theta$ and $\phi$ are the azimuth and elevation steering angles, respectively.

 The association matrix $\mathbf{C} \in \{0,1\}^{J \times (1+K)}$ is defined as:
\begin{equation}
\small
\mathbf{C}=\left[\mathbf{c}_{0}, \mathbf{c}_{1}, \mathbf{c}_{2}, \cdots, \mathbf{c}_{K}\right]=\left[\begin{array}{c}
{c_{10}\  c_{11}\  c_{12}   \cdots\  c_{1 K}} \\
\vdots \\
{c_{J 0}\  c_{J 1}\  c_{J 2} \cdots\  c_{J K}}
\end{array}\right],
\end{equation}
where \!$\mathbf{c}_{0} = [c_{10}, c_{20}, \cdots, c_{J0}]^{T}\in \{0,1\}^{J \times 1}$ and $\mathbf{c}_{k} = [c_{1k}, c_{2k}, \cdots, c_{Jk}]^{T} \in \{0,1\}^{J \times 1}$ represent the association vector of the RIS and UE-$k$, respectively.
Besides,  $c_{j0}$ and $ c_{jk}$ are binary decision variables, which determine the RIS and UE-$k$ associated with BS-$j$, respectively.
 Since RIS can only serve one BS and each UE can only be served by one BS, we have $\|\mathbf{c}_{0}\|_{0} = 1$ and $\|\mathbf{c}_{k}\|_{0} = 1$.

Then, the signal received by user-$k$ from BS-$j$ can be expressed as
\begin{small}
\begin{equation}
y_{j, k}=\widetilde{\mathbf{h}}_{j,k} \mathbf{w}_{j,k} s_{j,k}+\widetilde{\mathbf{h}}_{j,k} \sum_{i \neq k, i \in \mathcal{Q}_{j}} \mathbf{w}_{j,i} s_{j,i}+n_{k}, \ j \in \mathcal{J}, k \in \mathcal{K},
\end{equation}
\end{small}where $\widetilde{\mathbf{h}}_{j,k}=c_{jk}(\mathbf{h}_{\mathrm{d},j,k}^{H}+c_{j0}\mathbf{h}_{\mathrm{r},j,k}^{H}\mathbf{\Psi} \mathbf{G}_{j} )$ denotes the equivalent channel between BS-$j$ and UE-$k$, $\mathcal{Q}_{j}$ denotes the index set of users associated with BS-$j$, $\mathbf{w}_{j,k} \in \mathbb{C}^{N \times 1}$ denotes the beamforming vector of BS-$j$ for UE-$k$, $s_{j,k}$ denotes the data symbol transmitted independently from BS-$j$ to UE-$k$ with zero mean and unit power, and $n_{k} \sim \mathcal{CN}(0,\sigma^{2})$ denotes the complex additive white Gaussian noise at UE-$k$. Assuming that neighboring BSs are operated at different frequency bands and the intra-cell interference can be ignored. Accordingly, the signal-to-interference-plus-noise (SINR) of user-$k$ is written as:
 \begin{equation}
 \small
\gamma_{j,k}=\frac{\left|\widetilde{\mathbf{h}}_{j,k} \mathbf{w}_{j,k}\right|^{2}}{\sum_{i \neq k, i \in \mathcal{Q}_{j}}\left|\widetilde{\mathbf{h}}_{j,k} \mathbf{w}_{j,i}\right|^{2}+\sigma^{2}}, \ j \in \mathcal{J}, k \in \mathcal{K}.
\end{equation}
\vspace{-0.5 cm}
\subsection{Channel Model}
For the RIS-assisted BS-$j$, the channel of BS-RIS is modeled according to the extensively used 3D Saleh-Valenzuela channel model \cite{3Dchannel} as
 \begin{equation}
 \small
 \mathbf{G}_{j} = \sum_{\ell=0}^{L} \alpha_{\ell}
  \mathbf{a}_{\mathrm{M}}^{\mathrm{H}}(\phi_{\mathrm{AoA}}^{(\ell)}, \theta_{\mathrm{AoA}}^{(\ell)})\mathbf{a}_{\mathrm{N}}(\phi_{\mathrm{AoD}}^{(\ell)}),
 \end{equation}
where $L$ denotes the number of NLoS paths, and $ \alpha_{\ell}$ is the complex gain of the $\ell$-th path. Here, the parameter $\phi_{\mathrm{AoD}}^{(\ell)}$ represents the angle of departure (AoD) of the BS in the $\ell$-th path. Also, the parameters $\phi_{\mathrm{AoA}}^{(\ell)}$ and $\theta_{\mathrm{AoA}}^{(\ell)}$  represent the elevation and azimuth angles of arrival (AoA) for two-dimensional RIS in the $\ell$-th path, respectively. We employ a uniform linear array (ULA) at the BS. Thus, the array response of the BS can be expressed as
 \begin{equation}
 \small
 \mathbf{a}_{\mathrm{N}}(\phi_\mathrm{AoD}) = [1,e^{j\pi sin(\phi_\mathrm{AoD})},\cdots,e^{j\pi (N-1) sin(\phi_\mathrm{AoD})}],
  \end{equation}
and  $\mathbf{a}_{\mathrm{M}}  (\phi_\mathrm{ AoA},\theta_\mathrm{ AoA}) =  \mathbf{a}_{\mathrm{M}_{el}}( \phi_\mathrm{AoA} ) \otimes \mathbf{a}_{\mathrm{M}_{az}}( \theta_\mathrm{AoA} )$ is the array steering vector of RIS.
The array steering vectors $\mathbf{a}_{\mathrm{M}_{el}} \left( \phi_\mathrm{AoA} \right)$ and $\mathbf{a}_{\mathrm{M}_{az}}\left( \theta_\mathrm{AoA} \right)$ have the same definition model as
$\mathbf{a}_{\mathrm{N}}\left(\phi_\mathrm{AoA}\right)$.

Typically, the RIS is deployed on the surface of buildings around UEs, leading to a high probability of LoS path between RIS and UEs. Thus, the channel between the RIS and the $k$-th UE can be given by
\begin{equation}
\small
\label{h}
\mathbf{h}_{\mathrm{r},j,k} = \alpha_{k} \xi_{\mathrm{t}} \xi_{\mathrm{r}} \mathbf{a}_{\mathrm{M}} \left( \phi_{\mathrm{AoD}}, \theta_{\mathrm{AoD}} \right)
\end{equation}
where $\alpha_{k}$ is the channel gain, and $\xi_{\mathrm{t}}$ and $\xi_{\mathrm{r}}$ are the transmit and receive antenna gains, respectively. Besides, BS-UE channels $\mathbf{h}_{\mathrm{d},k,j}$ can also be obtained according to (\ref{h}).
\vspace{-0.3 cm}
\subsection{Problem Formulation}
We define the cascaded channel and beamforming matrix of BS-$j$ as $\mathbf{H}_{j} \triangleq [ \widetilde{\mathbf{h}}_{j,1}^{H},\cdots, \widetilde{\mathbf{h}}_{j,|\mathcal{Q}_{j}|}^{H}] \in \mathbb{C}^{N \times |\mathcal{Q}_{j}|}$ and
$\mathbf{W}_{j} \triangleq [ \mathbf{w}_{j,1},\cdots, \mathbf{w}_{j,|\mathcal{Q}_{j}|}] \in \mathbb{C}^{N \times |\mathcal{Q}_{j}|}$, respectively. In this paper, we aim to maximize the downlink sum-rate of all users by  jointly optimizing the beamforming and the BS-RIS-UE association with the constraints of transmit power of each BS and the quality of service (QoS) of each UE. Accordingly, the sum-rate maximization problem can be formulated as
 \begin{subequations}
 \small
\begin{align}
&\max _{\mathbf{W},\mathbf{\Psi},\mathbf{C}}  \ \  \  \sum_{k \in \mathcal{K}} \sum_{j \in \mathcal{J}}  \log _{2}\left(1+\gamma_{j,k}\right)\\
&\text { s.t. }  \quad \phi,\theta \in \mathcal{F},  \\
&\quad\quad\quad  R_{k}=\sum_{j \in \mathcal{J}}  \log _{2}\left(1+\gamma_{j,k}\right) \geq R_{\mathrm{min}}, \forall  k \in \mathcal{K},\\
& \quad\quad\quad\sum_{k \in Q_{j}}\left\|{\mathbf{w}_{j,k}}\right\|^{2} \leq P_{j}, \forall j \in \mathcal{J}, \\
&\quad\quad\quad \sum_{j \in \mathcal{J}} c_{j, k}=1, \forall k \in \mathcal{K}, \\
&\quad\quad\quad \sum_{k \in \mathcal{K}} c_{j, k} \geq 1, \forall j \in \mathcal{J},\\
& \quad\quad\quad\sum_{j \in \mathcal{J}} c_{j, 0} = 1,
\end{align}
\end{subequations}
where
$\mathcal{F}=\left\{0, \frac{2 \pi}{2^{B}}, \frac{2 \pi \times 2}{2^{B}}, \cdots, \frac{2 \pi \times\left(2^{B}-1\right)}{2^{B}}\right\}$ is the feasible set for $\phi$ and $\theta$. Additionally, (8c) satisfies the minimum communication rate of each UE, (8d) bears that BS-$j$
 satisfies the transmission power bridge, (8e) ensures that each UE is only served by one BS, (8f) ensures that RIS only serves one BS at each coherent period, and (8g) allows each BS to serve multiple UEs at the same time. To simplify the problem, we consider zero forcing (ZF) precoding at BS-$j$ with satisfying power constraint $\mathbf{W}_{j}^{*} = \sqrt{P_{j}} \mathbf{H}_{j}^{+}$ \cite{ZF}.
However, the above optimization problem is still an NP-hard problem and is difficult to solve by the conventional optimization methods due to the non-convex mixed-integer constraint.
Moreover, the computational complexity of traditional algorithms will also increase dramatically with the complex multi-BS and multi-user communication environment. Therefore, we adopt DRL algorithm to solve this mathematically difficult non-linear and non-convex problem, which will be presented in the next section.

\section{SAC-DRL Based Joint Beamforming and Association  Design}
In this section, we design the passive beamforming and BS-RIS-UE association jointly based on SAC-DRL algorithm. The problem is formulated as a Markov Decision Process (MDP) model with unknown status transition function.  The MDP model and SAC-DRL algorithm  are described as follows.
\vspace{-0.3 cm}
\subsection{MDP Model}
Reinforcement learning (RL) is a branch of machine learning without prior information. It is a model of trial-and-error learning that continues to interact with the environment through agent. According to the feedback of the environment, action in turn changes the state of the environment to obtain the maximum long-term cumulative rewards. In this paper, beamforming and BS-RIS-UE association can be modeled as an MDP in the same form as the sequential decision-making problem.
Particularly, we formulate the MDP model contains five elements, $<\mathcal{S},\mathcal{A},\mathcal{P},\mathcal{R},\gamma>$, where $\mathcal{S}$ denotes a finite set of the environment states, $\mathcal{A}$ denotes a finite set of actions, $\mathcal{P}$ denotes a finite set of transition probabilities, which means that the agent takes action $a$ with probability $p$ to transfer to state $s'$ in the state $s$, $\mathcal{R}$ denotes a finite set of immediate reward and $\gamma$ denotes the discount factor of reward.
The detailed tuple in our proposed model is expressed as follows:
\begin{itemize}
\item Action ${a}_{t}$: The actions are the optimal variables, including azimuth angle $\theta$, elevation angle $\phi$, and varibles associated with RIS and UEs. Thus, the action can be expressed as
\begin{equation}
{a}_{t}=\left\{\theta^{({t})}, \phi^{(t)},\{c_{j0}^{(t)}\}_{j \in \mathcal{J}}, \{c_{jk}^{(t)}\}_{j \in \mathcal{J}, k \in \mathcal{K}}\right\}.
\end{equation}
\item State ${s}_{t}$: The state of the system mainly consists of two parts, i.e., the rate of each UE and the cascaded channel of each BS. Then, we define the state of the $t$-th step as
\begin{equation}
{s}_{t}=\left\{ \{R_{k}^{(t)}  \}_{k \in \mathcal{K}}, \{ \mathbf{H}_{j}^{(t)} \}_{j \in \mathcal{J}}   \right  \}.
\end{equation}
\item Reward ${r}_{t}$: The objective is to maximize the achievable rate. Therefore, we set the reward function equivalent to the objective equation:
    \begin{equation}
  {r}_{t} =  \sum_{k \in \mathcal{K}} \sum_{j \in \mathcal{J}}  \log _{2}\left(1+\gamma_{j,k}\right).
    \end{equation}
\end{itemize}
\vspace{-0.4 cm}
\subsection{Fundament of SAC-DRL Algorithm}
The goal of traditional RL is to find the optimal policy to maximize the expectation of cumulative reward, namely
\begin{equation}
\small
\pi^* = \mathrm{arg} \max_{\pi}\mathbb{E}_{({s}_{t},{a}_{t})\sim \rho_{\pi}} \left[\sum_{t}r({s}_{t}, {a}_{t})\right],
\end{equation}
where $\pi$ denotes a policy that mapping from a state to an action, $\rho_{\pi}$ denotes the state-action trajectory distribution formed by the policy $\pi$, ${s}_{t}$ and ${a}_{t}$ denote the state and action at step $t$, respectively.

In this paper, we utilize SAC algorithm, which combines the maximum entropy framework in the reward function \cite{haarnoja2018soft}.
The goal of SAC is to obtain the maximum entropy of the action distribution in the current state while obtaining the maximum cumulative reward. Thus, the objective function in SAC algorithm is defined as
\begin{equation}
\small
\pi^* = \mathrm{arg} \max_{\pi}\mathbb{E}_{({s}_{t},{a}_{t})\sim \rho_{\pi}} \left[\sum_{t}r({s}_{t}, {a}_{t})+\alpha \mathcal{H}(\pi(\cdot\mid {s}_{t})) \right],
\end{equation}
where $\alpha$ is the adjustable temperature parameter that adjusts the tradeoff between reward and entropy term, $\mathcal{H}(\pi(\cdot\mid {s}_{t}))$ is the information  entropy of policy $\pi$ which is defined as $\mathcal{H}(\pi(\cdot \mid {s}_{t}))\triangleq - \mathrm{log} (\pi(\cdot \mid {s}_{t}))$. It is worth pointing out that under certain state, maximizing entropy can increase the randomness of policy selection at the beginning of training to find the optimal strategy, which is also the nature of SAC to enhance the random exploration of agent.

In the SAC algorithm, the soft policy iteration achieves the maximum goal  by alternating policy evaluation and policy improvement.
In policy evaluation,
 the soft Q-function according to the Bellman iteration function can be computed by
\begin{equation}
Q({s}_{t},{a}_{t}) = r_{t} + \gamma \mathbb{E}_{{s}_{t+1}\sim p}[{V}({s}_{t+1})],
\end{equation}
where ${V}({s}_{t+1})$
 is the soft value function and is given by
\begin{equation}
 {V}({s}_{t+1}) =\mathbb{E}_{{a}_{t+1} \sim \pi}[Q({s}_{t+1}, {a}_{t+1})-\alpha \log \pi({a}_{t+1} \mid {s}_{t+1})].
\end{equation}
Under the large continuous domain setting, we need a neural network to parameterize the soft Q-function $Q_{\boldsymbol{\theta}}({s}_{t},{a}_{t})$
with a parameter $\boldsymbol{\theta}$.
Then, the parameters are trained to minimize the soft Bellman residual
\begin{equation}
\begin{aligned}
J_{Q}(\boldsymbol{\theta})=\mathbb{E}_{({s}_{t}, {a}_{t}) \sim \mathcal{D}}\bigg[\frac{1}{2}(Q_{\boldsymbol{\theta}}({s}_{t}, {a}_{t})-(r({s}_{t}, {a}_{t}) \\
+\gamma \mathbb{E}_{{s}_{t+1} \sim p}\left[V_{\bar{\boldsymbol{\theta}}}({s}_{t+1})\right]))^{2}\bigg],\label{eq:Q}
\end{aligned}
\end{equation}
where $\mathcal{D}$ denotes the experience replay memory and $V_{\bar{\boldsymbol{\theta}}}({s}_{t+1})$ denotes the soft value estimated using a target network.
\begin{figure}[t]
\centering
\includegraphics[width = 2.8 in]{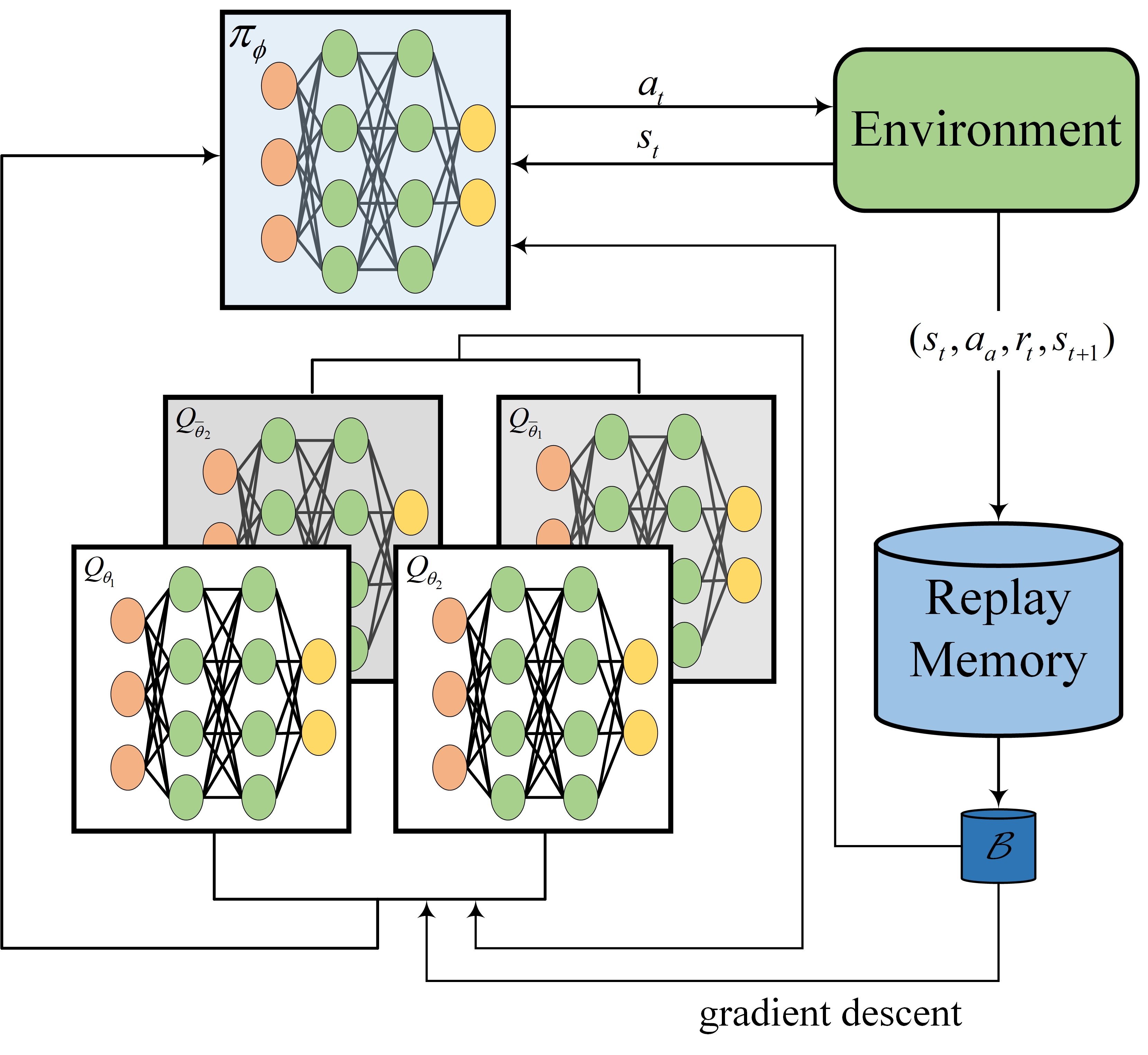}
\vspace{-0.1 cm}
\caption{The architecture of the SAC-DRL algorithm.}
\vspace{-1 cm}
\end{figure}
In the policy improvement step, the new policy updates in the exponential direction of the Q-function. For simplicity, we update the policy according to Kullback-Leibler (KL) divergence. A neural network is also required to parameterize the policy function with a parameter $\boldsymbol{\phi}$. Then, the parameter is trained to minimize the expected KL divergence as
\begin{equation}
\small
J_{\pi}(\boldsymbol{\phi})=\mathbb{E}_{{s}_{t}\sim \mathcal{D}}\bigg[ D_\mathrm{K L}\bigg(\pi_{\boldsymbol{\phi}}(\cdot \mid {s}_{t}) \bigg \| \frac{\exp (Q_{\boldsymbol{\theta}}({s}_{t}, \cdot))}{Z_{\boldsymbol{\theta}}({s}_{t})}\bigg)\bigg], \label{policy}
\end{equation}
where $Z_{\boldsymbol{\theta}}({s}_{t})$  is adopted to normalize the distribution.

Besides, the temperature parameter $\alpha$ is tuned by an automation entropy adjustment method as \cite{ASAC}
\begin{equation}
J(\alpha) = \mathbb{E}_{{a}_{t}\sim \pi_{\boldsymbol{\phi}}}[-\alpha \mathrm{log} \pi_{\boldsymbol{\phi}}({a}_{t}|{s}_{t})-\alpha \bar{\mathcal{H}}], \label{alpha}
\end{equation}
where $\bar{\mathcal{H}}$ is the minimum expected target entropy.
\vspace{-0.3 cm}
\subsection{The Architecture of SAC-DRL Algorithm}
The architecture of SAC-DRL algorithm for designing passive beamforming and BS-RIS-UE association problem is shown in Fig. 2.
In the framework, there is an actor policy network $\pi_{\boldsymbol{\phi}}$ that can interact with the environment and generate an action distribution (${a}_{t} \sim \pi_{\boldsymbol{\phi}}({a}_{t} \mid {s}_{t})$) in the current state. There are also two critic networks $Q_{\boldsymbol{\theta}_{1}}$, $Q_{\boldsymbol{\theta}_{2}}$ to evaluate the quality of action ${a}_{t}$ in state ${s}_{t}$ based on the Bellman equation. Then, two target critic networks $Q_{\bar{\boldsymbol{\theta}}_{1}}$, $Q_{\bar{\boldsymbol{\theta}}_{2}}$ are established to avoid the overestimation of Q-value. Besides, an experience replay buffer is used to store tuples in the MDP model.

During each training episode, firstly, a state ${s}_{0}$ is initialized randomly. Secondly, in this state, the agent samples the control action ${a}_{t}$ from the actor policy network and obtains the reward ${r}_{t}$ at this moment, and then the state is transferred to the next state ${s}_{t+1}$. Next, we store ${s}_{t}$, ${a}_{t}$, ${r}_{t}$, and ${s}_{t+1}$ as a sample in the experience replay memory $\mathcal{D}$. Once the $\mathcal{D}$ is full, the new sample will replace the oldest sample. Finally, we extract a small batch $\mathcal{B}$ of samples from $\mathcal{D}$ to train networks and use the stochastic gradient method to update networks.
At the end of each training episode, we can find the optimal policy from the updated actor policy network, that is, the optimal
passive beamforming $\mathbf{\Psi}(\phi,\theta)$ and BS-RIS-UE association matrix $\mathbf{C}$ in this episode.

The complete procedure of SAC-DRL based jointly design of passive beamforming and BS-RIS-UE association is summarized as Algorithm I.
\begin{algorithm}[t]
\small
\caption{SAC-DRL based Passive Beamforming and BS-RIS-UE Association Design}
\label{alg:1}
\begin{algorithmic}[1]

\STATE{Randomly initialize network parameters $\boldsymbol{\theta}_{1}$, $\boldsymbol{\theta}_{2}$, $\boldsymbol{\phi}$.}
\STATE{Initialize the target network parameters $\bar{\boldsymbol{\theta}}_{1} \leftarrow\boldsymbol{\theta}_{1}$, $\bar{\boldsymbol{\theta}}_{2} \leftarrow\boldsymbol{\theta}_{2}$.}
\STATE{Initialize experience replay memory $\mathcal{D}$.}

            \FOR {each episode }
                \STATE{Initialize state ${s}_{0} \in \mathcal{S}, {s} \leftarrow {s}_{0}$.}
                \FOR {each environment step }
                \STATE{${a}_{t} \sim \pi_{\boldsymbol{\phi}}\left({a}_{t} \mid {s}_{t}\right)$;}
                \STATE{${s}_{t+1} \sim p\left({s}_{t+1} \mid {s}_{t}, {a}_{t}\right)$;}
                \STATE{$\mathcal{D} \leftarrow \mathcal{D} \cup\left\{\left({s}_{t}, {a}_{t}, {r}_{t}, {s}_{t+1}\right)\right\}$.}
                \ENDFOR
                  \FOR {each gradient step }
                   \STATE{Randomly select a sample batch from memory;}
                   \STATE{Update Q-function parameters with (\ref{eq:Q});}
                   \STATE{$\boldsymbol{\theta}_{i} \leftarrow \boldsymbol{\theta}_{i}-\lambda \nabla_{\boldsymbol{\theta}_{i}} J_{Q}\left(\boldsymbol{\theta}_{i}\right)$ for $i \in\{1,2\}$;}
                    \STATE{Update policy function parameter with (\ref{policy});}
                    \STATE{$\boldsymbol{\phi} \leftarrow \boldsymbol{\phi}-\lambda \nabla_{\boldsymbol{\phi}} J_{\pi}(\boldsymbol{\phi})$ ;}
                     \STATE{Update  temperature coefficient with (\ref{alpha});}
                     \STATE{$\alpha \leftarrow \alpha-\lambda \nabla_{\alpha} J_{\alpha}(\alpha) $;}
                     \STATE{Update target network;}
                      \STATE{$\bar{\boldsymbol{\theta}}_{i} \leftarrow \tau \boldsymbol{\theta}_{i}+(1-\tau) \bar{\boldsymbol{\theta}}_{i}$ for $i \in\{1,2\}$.}
                  \ENDFOR
            \ENDFOR
   \end{algorithmic}
\end{algorithm}

\section{Simulation Results}
In this section, we present simulation results of our proposed SAC-DRL based joint beamforming and BS-RIS-UE association design for the RIS-assisted mmWave system. We assume three BSs located in $(0m,0m)$, $(200m,0m)$ and $(0m,200m)$, respectively, jointly serve $K = 16$ single-antenna UEs which are randomly distributed in a circular area at $(150m, 100m)$ with a radius of $30m$. The RIS has $M = 64$  passive elements and is deployed at $(50m,100m)$.
In the mmWave channel model, the NLOS complex gain $\alpha_{\ell}$ is set as $\alpha_{\ell}\sim\mathcal{ C}\mathcal{N}(0,10^{-0.1\kappa})$ where $\kappa =\kappa_{a}+10\kappa_{b}log_{10}(d)+\kappa_{c} $ with $\kappa_{a} = 72$, $\kappa_{b} = 2.92$, $\kappa_{c}\sim\mathcal{ C}\mathcal{N}(0,\sigma_{c}^{2})$ and $\sigma_{c} = 8.7 $ dB. The LOS complex gain $\alpha_{k}$ is set the same as $\alpha_{\ell}$, while other parameters are set as $\kappa_{a}' = 61.4$, $\kappa_{b}' = 2$ and $\sigma_{c}' = 5.8 $ dB. Other required parameter settings are as follows: $\sigma^{2} = -85$ dBm, $P_{\mathrm{max}} = 30$ dBm, $L = 5$, $\xi_{t} = 9.82$ dBi and $\xi_{r}=0$ dBi.
 In addition, the hyperparameters of the proposed SAC algorithm is summarized in Table I.
    \begin{table}[t]
   \small
		\centering
        \caption{SAC hyperparameters}
		\begin{tabular}{l l}
        \hline
        Hyperparameter&Value\\\hline
		Layers&	2 fully connected layers\\
		Layer hidden units	&  256\\
        Activation function  &  ReLU\\
        Batch size  &  64\\
        Replay buffer size  &  $10^6$\\
        Target smoothing coefficient  &   0.005\\
        Learning rate  & 0.0001 \\
        Discount rate  & 0.95 \\
       Target update interval & 1\\
       Gradient steps  & 1 \\
        Loss  & Mean squared error  \\
        \hline
		\end{tabular}
		\label{tab:Margin_settings}
	\end{table}
\begin{figure}[t]
\centering
\includegraphics[width = 3 in]{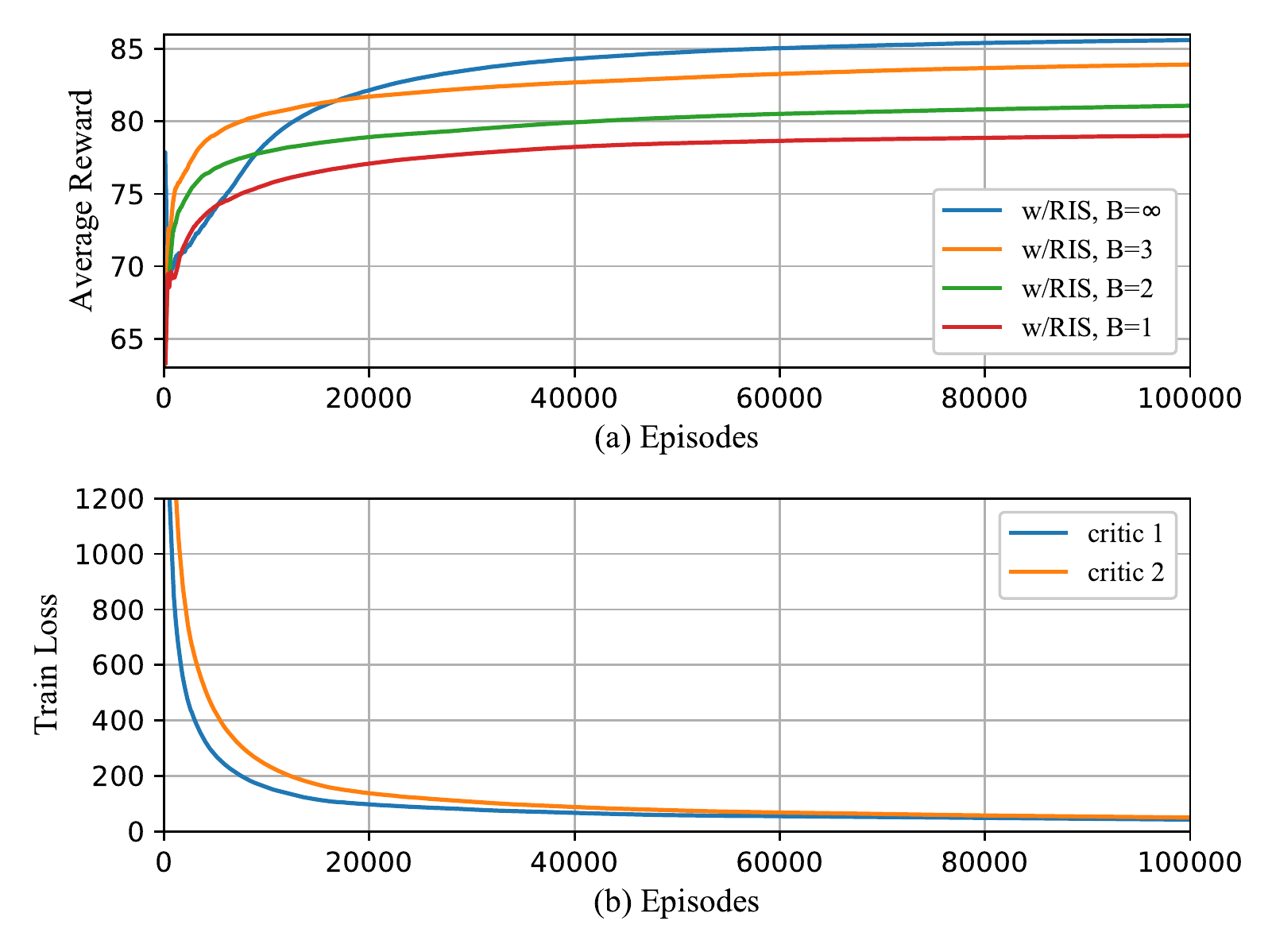}
\caption{Convergence performance of the proposed algorithm.}
\label{fig:C}
\vspace{-0.25 cm}
\end{figure}

\begin{figure}[t]
\centering
\includegraphics[width = 3 in]{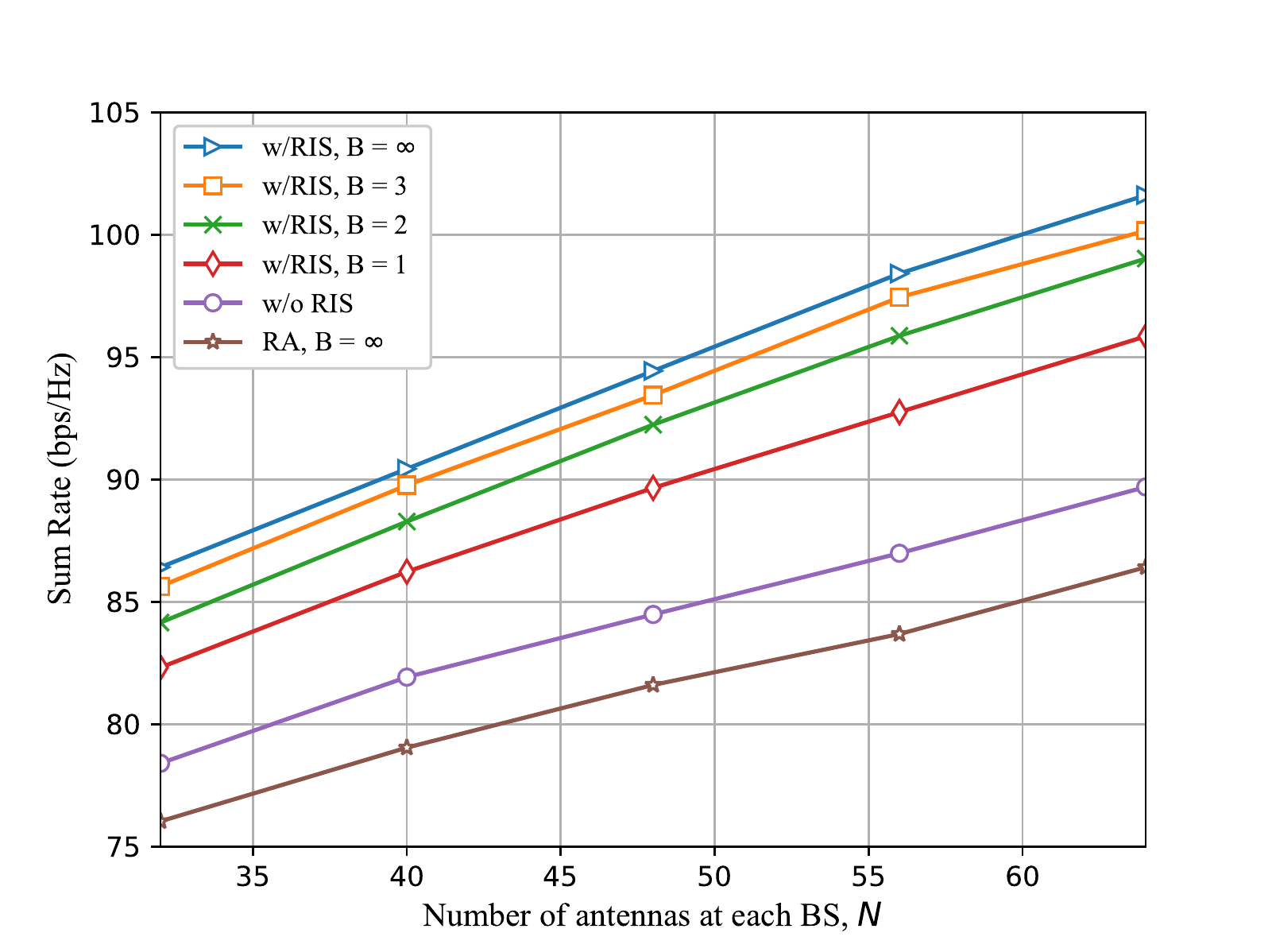}
\caption{Sum-rate versus $N$ with $K$=16, $M$=64.}
\label{fig:N}
\end{figure}

\begin{figure}[t]
\centering
\includegraphics[width = 3 in]{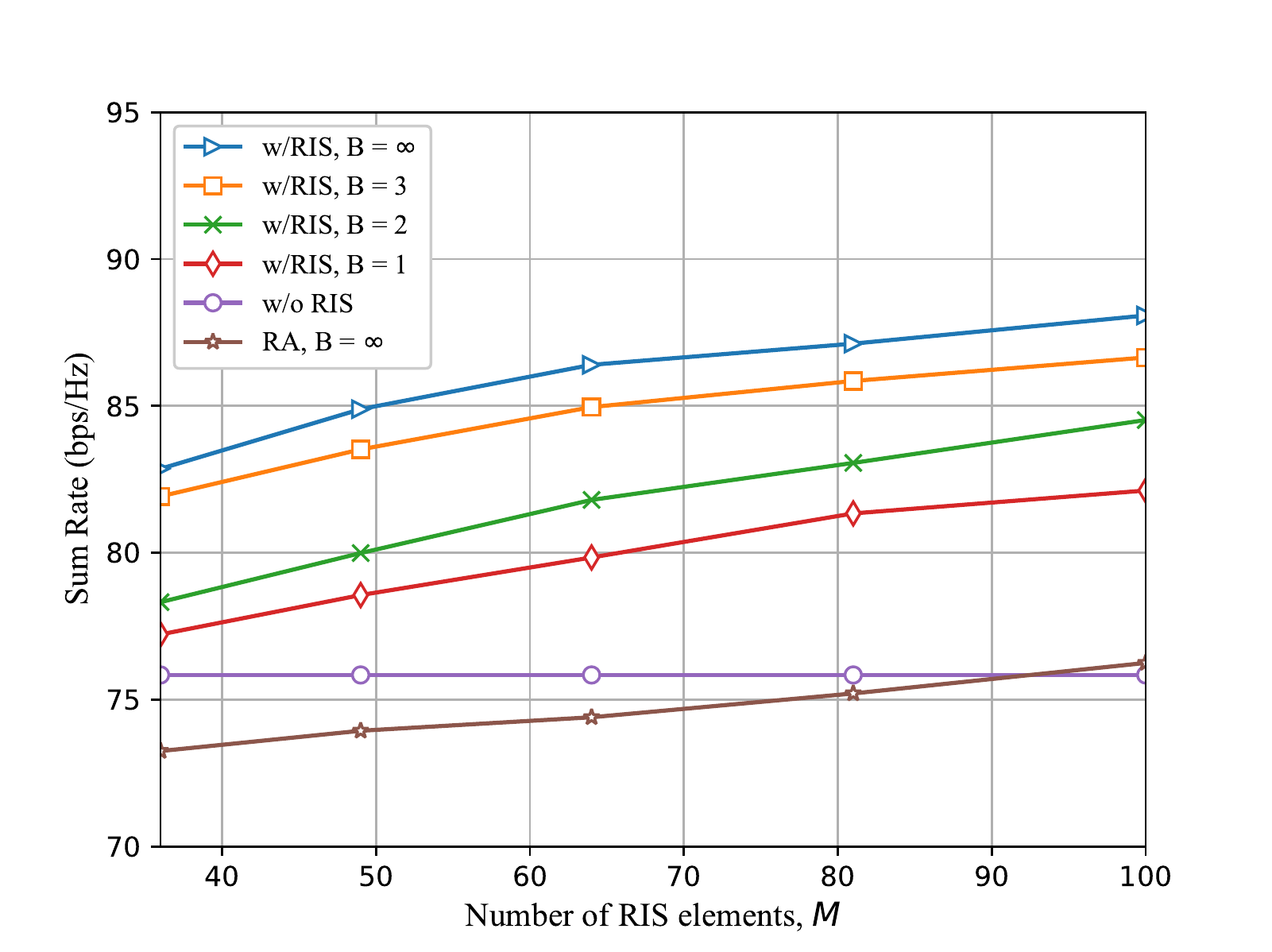}
\caption{Sum-rate versus $M$ with $K$=16, $N$=32.}
\label{fig:M}
\end{figure}

\begin{figure}[t]
\centering
\includegraphics[width = 3 in]{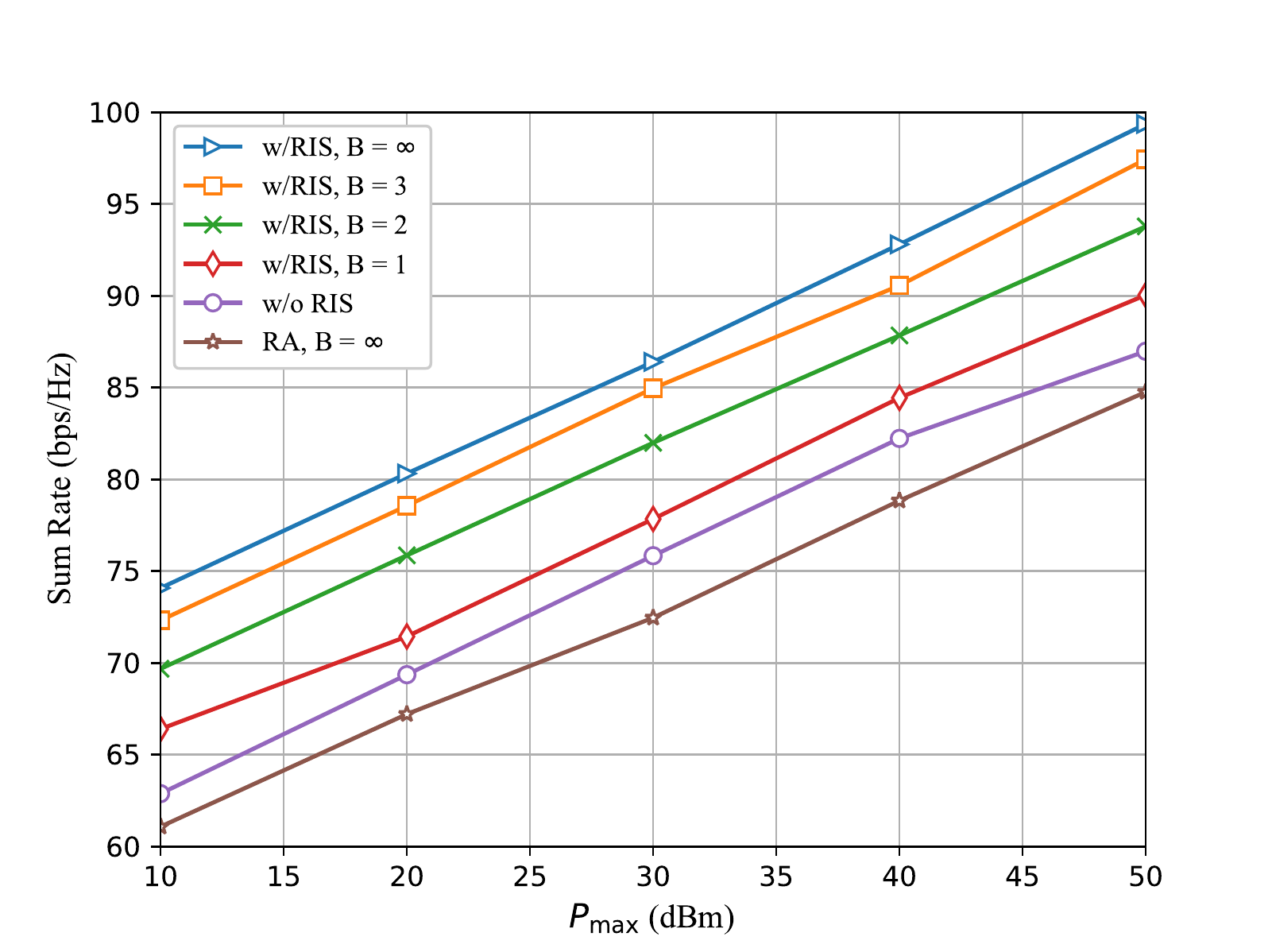}
\caption{Sum-rate versus the maximum transmit power.}
\label{fig:PT}
\vspace{-0.3 cm}
\end{figure}

\begin{figure}[t]
\centering
\includegraphics[width = 3 in]{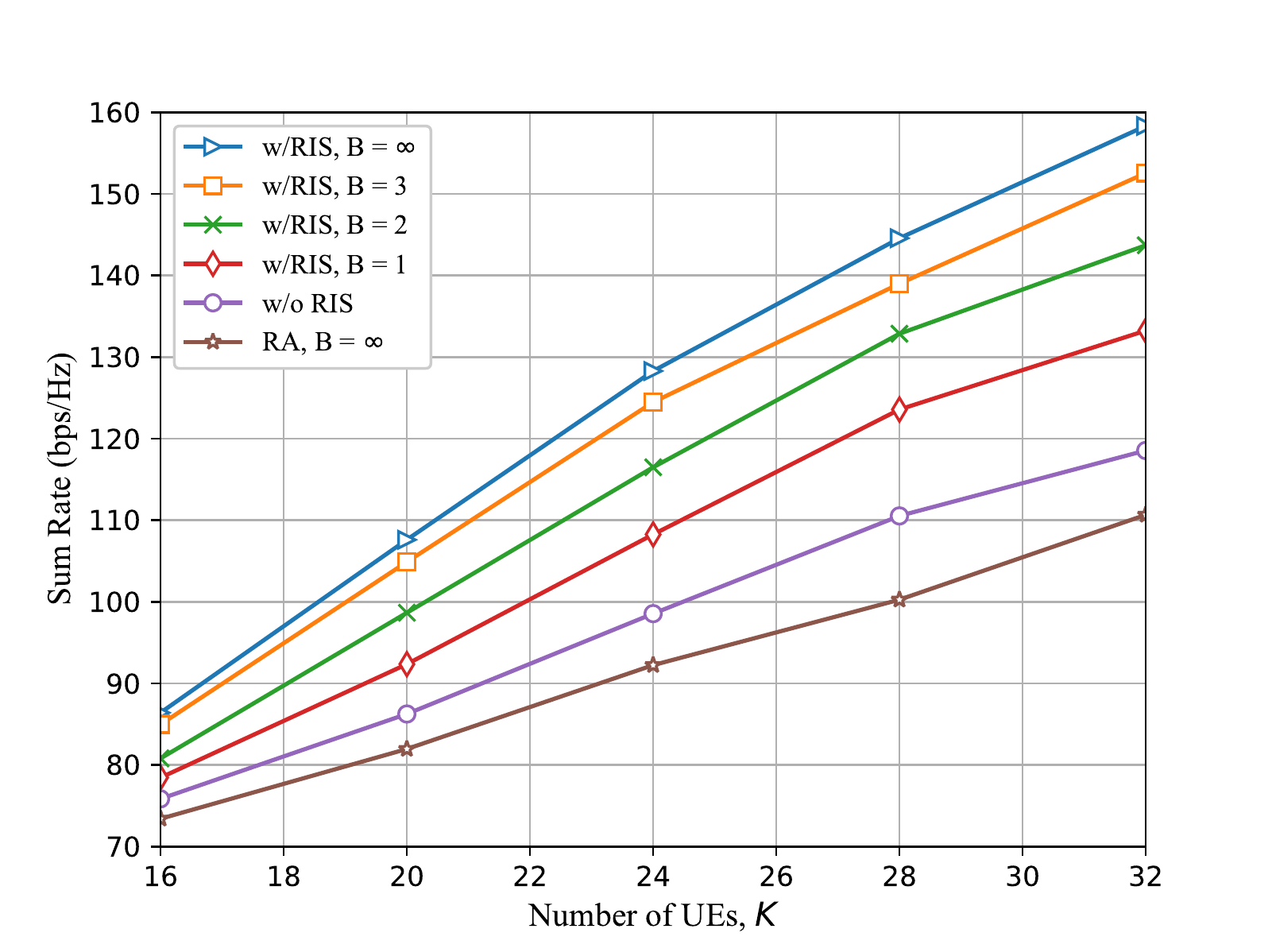}
\caption{Sum-rate versus the number of UEs.}
\label{fig:K}
\vspace{-0.4 cm}
\end{figure}

\begin{figure}[t]
\centering
\includegraphics[width = 2.9 in]{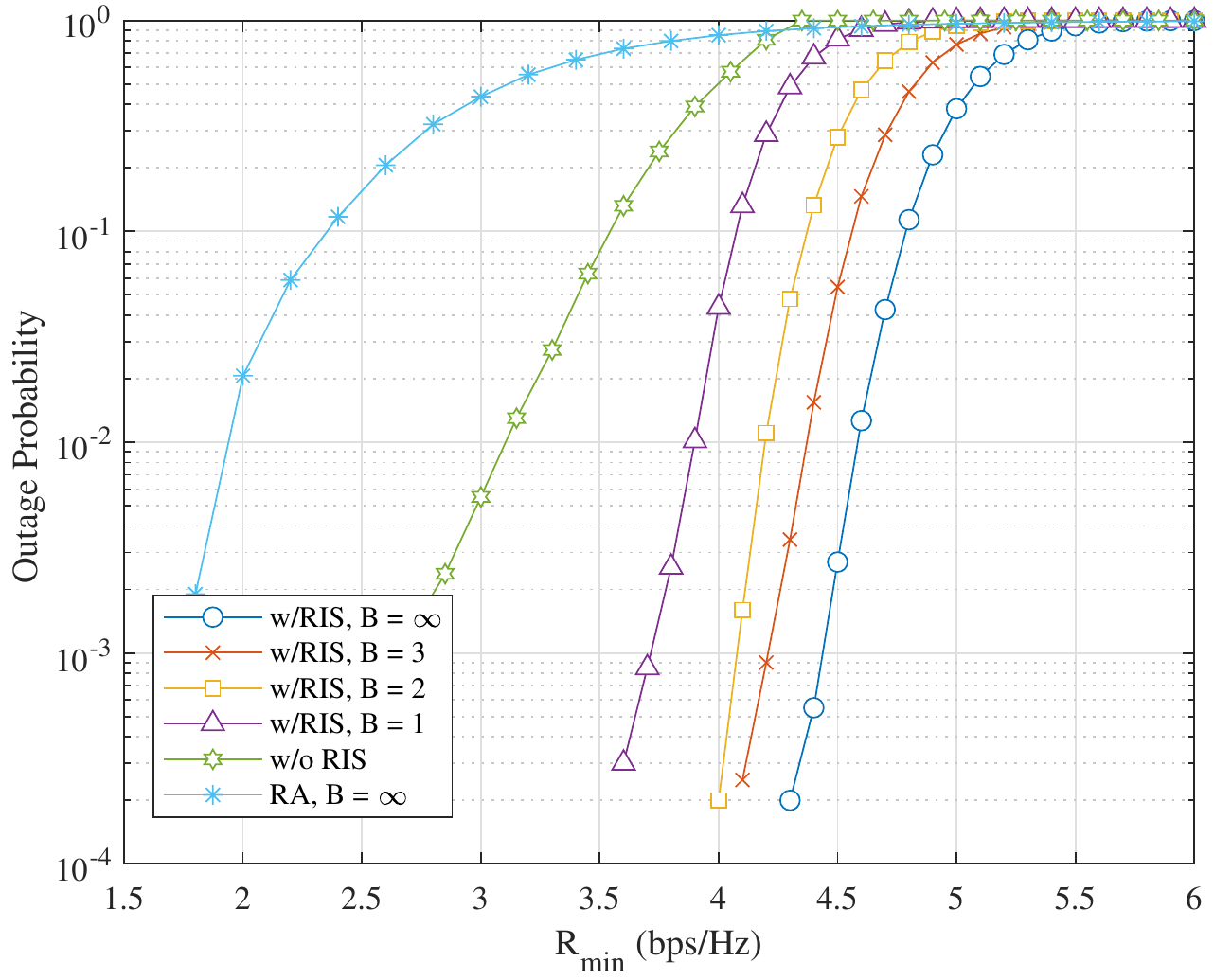}
\caption{Outage probability versus $R_{\mathrm{min}}$.}
\label{fig:pout}
\vspace{-0.6 cm}
\end{figure}

Fig. \ref{fig:C} shows the convergence performance of the proposed algorithm under different resolution quantization codebook. The average reward versus learning episodes is shown in Fig. \ref{fig:C}(a).
In order to render the convergence curve more aesthetic, we plot the average reward in this figure. Simulation results illustrate that all curves can converge to a steady state. As the number of quantization bits $B$ increases, the convergence rate slows down but obtains significant improvement on average reward.
Fig. \ref{fig:C}(b) illustrates the train loss value versus the number of episodes of the case of continuous phase-shift (i.e. $B=\infty$). It can be seen that with the increase of training times, the training loss of two critic networks will decrease rapidly and reach a stable state. From this point of view, the convergence of our proposed SAC-DRL algorithm is also illustrated.

Fig. \ref{fig:N} presents sum-rate versus the number of antennas at each BS.
 For comparison purposes, we also include two benchmarks: $i)$ Scheme without RIS (denoted as "w/o RIS") and BS-UE association is optimized; $ii)$ Scheme with RIS but BS-RIS-UE association is randomly associated (denoted as "RA").
 We can observe that the sum-rate increases with the number of antennas. Compared with the two benchmarks, the proposed algorithm can achieve much higher performance, which verifies the importance of BS-RIS-UE association and the effectiveness of the proposed DRL-based algorithm.

Fig. \ref{fig:M} describes the sum-rate versus the number of RIS elements. It is obvious that sum-rate increases with the number of RIS elements. In addition, compared with benchmarks, the proposed algorithm can achieve better performance with the improvement of quantization accuracy. Fig. \ref{fig:PT} presents the sum-rate versus different settings of $P_{\mathrm{max}}$ and similar conclusions can be obtained.

Fig. \ref{fig:K} shows the sum-rate versus the number of UEs. From the Fig. \ref{fig:K}, we can see that with the increase of the number of UEs, our proposed algorithm outperforms other comparison schemes, especially when $K>24$, our algorithm can still achieve competitive performance.

Fig. \ref{fig:pout} illustrates the impact of the minimum communication rate of each user on the outage probability. The outage probability is calculated as $P_{\mathrm{out}}(R_{\mathrm{min}}) = \mathrm{Pr}(\mathbb{E}[R_{k}]\leq R_{\mathrm{min}})$.
We can see that the outage probability increases with the increase of the rate, and the outage probability of our algorithm is greatly reduced. When $R_{\mathrm{min}} = 5$bps/Hz, the outage probabilities of other algorithms are close to 1 except for the case of continuous phase-shift (i.e. $B=\infty$), which illustrates the reliability improvement of the RIS-assisted mmWave system.

\vspace{-0.2 cm}
\section{Conclusions}
In this paper, we considered a RIS-assisted multi-BS multi-user mmWave communication system and investigated sum-rate maximization problem by jointly optimizing passive beamformer and BS-RIS-UE association. To avoid the extremely high computational complexity caused by complex communication scenarios, we propose a SAC-DRL algorithm to joint design  beamformer  and BS-RIS-UE association under the condition of non-convex constrains.
The experimental results demonstrated that our proposed SAC-DRL algorithm can achieve better performance compared with benchmarks.

\end{document}